\documentclass{PoS}

\usepackage[utf8x]{inputenc}

\usepackage{setspace}

\usepackage{physics}

\usepackage{graphicx}
\usepackage{wrapfig}
\graphicspath{{./images/}}

\title{Medium modified leading hadrons, jets and sub-jets in a single formalism}

\ShortTitle{Medium modified leading hadrons, jets and sub-jets in a single formalism}

\author{\speaker{Chathuranga Sirimanna}, Shanshan Cao and Abhijit Majumder
        \\Department of Physics and Astronomy, Wayne State University, Detroit, Michigan 48201, USA\\
        E-mail: \email{chathuranga.sirimanna@wayne.edu},  \email{shanshan.cao@wayne.edu}, \email{majumder@wayne.edu}}

\abstract{Vacuum and medium modified fragmentation functions are used to study the properties of hadrons produced in hard interactions, in $p$-$p$ and $A$-$A$ collisions, respectively. We study the modification of reconstructed jets and high transverse momentum (high-$p_\mathrm{T}$) hadrons as they propagate through a dense medium. Jets and jet modification are studied within the framework of a factorized jet function. The vacuum jet function is found consistent with PYTHIA simulations. Both the modification of the fragmentation function and the jet function are carried out within the identical higher-twist energy loss formalism. This approach is then extended to di-hadron production within a jet. Its further extension to a ``di-prong'' function and its application to the soft-drop measurements are discussed. The methodology of extending these semi-analytical calculations to full Monte-Carlo simulations is outlined. All calculations are carried out within a (2+1)-D viscous hydrodynamic simulation and compared to experimental data from a variety of collision energies. }

\FullConference{International Conference on Hard and Electromagnetic Probes of High-Energy Nuclear Collisions\\
		30 September - 5 October 2018\\
		Aix-Les-Bains, Savoie, France}

\begin{document}

\section{Introduction}
\label{Introduction}
\vspace{-10pt}
Jets, collimated sprays of particles, are produced in the early stage of relativistic heavy-ion collisions. They gather information about the Quark-Gluon Plasma (QGP) as they traverse it \cite{Majumder:2010qh}. Usually, jets are defined according to their energy and radius. Small radius jets are widely studied in heavy ion collisions which are expected to have little contamination from the QGP background. A jet function is usually defined to study small radius jets in theoretical calculations \cite{Dasgupta:2014yra,Kang:2016mcy}. Dasgupta et al. (DDSS) \cite{Dasgupta:2014yra} assumed the jet function at $R=1$ to be a delta function and performed an angular evolution down to smaller radii. Kang et al. (KRV) \cite{Kang:2016mcy} calculated the jet function at one scale using the “Soft Collinear Effective Field Theory” (SCET) \cite{Bauer:2000yr} and evolved it to different scales by using “Dokshitzer-Gribov-Lipatov-Altarelli-Parisi" (DGLAP) \cite{Dokshitzer:1977sg} evolution. Also, there has been considerable interest in studying jet substructure in the last decade \cite{Abdesselam:2010pt,Chang:2017gkt}, especially “Soft Drop”, a widely used jet substructure technique \cite{Larkoski:2014wba}. Even though there have been many theoretical studies on both small radius jets and “Soft Drop”, none have provided a common formalism to describe medium modified jets and “Soft Drop” together with leading hadrons.

In earlier work, leading hadrons with medium modified fragmentation functions have been studied in the “higher-twist” formalism \cite{Wang:2001ifa,Majumder:2007hx,Majumder:2004wh,Majumder:2004br}. In this current effort, we extended the evolution of the single hadron fragmentation function \cite{Wang:2001ifa,Majumder:2007hx} to the single jet function and succeeded in describing the small radius medium modified jets. The jet function was first calculated at lower energy and mass scale and then boosted to a higher energy and evolved to a higher scale. Di-hadron fragmentation function evolution \cite{Majumder:2004wh,Majumder:2004br} was reformulated to evolve the “di-prong function” in order to describe the soft dropped jet substructure. This calculation allows using a single formalism to describe medium modified hadrons, jets as well as jet substructure in a soft drop algorithm.

\section{Single hadron fragmentation function and single jet function}
\vspace{-10pt}
Although the vacuum evolution of single hadron fragmentation function through DGLAP evolution is similar in most studies, medium evolution is different in different formalisms. The quark jet function is defined as the probability to find a jet with momentum fraction $z$, energy $E_J$ and radius $R$ within the fragmentation products of a quark with energy $E_J/z$, and the operator definition is given by,
\begin{equation}
J_q(z, E_J, R, Q)=\frac{z^3}{2}\int \frac{d^4y d^4q e^{-iqy} \delta \big (z - \frac{p_J}{q^-}\big )}{(2 \pi)^4} Tr\bigg [\frac{\gamma^-}{2}  \mel {0}{\psi (0)}{p_J, S} \mel {p_J, S}{\bar{\psi} (y^+)}{0}\bigg ].
\label{SJF}
\end{equation}
Here, $Q=E_J R$ and sum over $S$ is assumed. The quark and gluon vacuum jet functions at lower energy ($E=10$~GeV) and large radius $(R=1.0)$ were calculated from PYTHIA simulation in lieu of experimental data. Then they were boosted up to a higher energy ($E=100$~GeV). Jets with high transverse momentum become collinear as they are boosted to higher energy. Since the virtuality, $Q \approx ER$, and the momentum fraction, $z$, are boost invariant, the boost shrinks the jet radius as the energy increases. For example, if we boost a jet function from $E=10$~GeV to $E=100$~GeV, the jet radius reduces by a factor of $10$. Finally, the boosted jet functions are evolved in scale up to a higher scale ($Q=10$~GeV to $Q=100$~GeV) using the DGLAP equations,
\begin{equation}
\frac{dJ_i\left(z, E_J, R, Q\right)}{d\log Q^2} = \frac{\alpha _ s}{2\pi} \int_z^1 \frac{dz\prime}{z\prime} P_{i\rightarrow j} \left(z\prime \right) J_j\left(\frac{z}{z\prime}, E_J, R, Q\right).
\label{Vac_Jet_evol}
\end{equation}

\begin{wrapfigure}{l}{0.5\textwidth}
    \centering
    \includegraphics[width=0.5\textwidth]{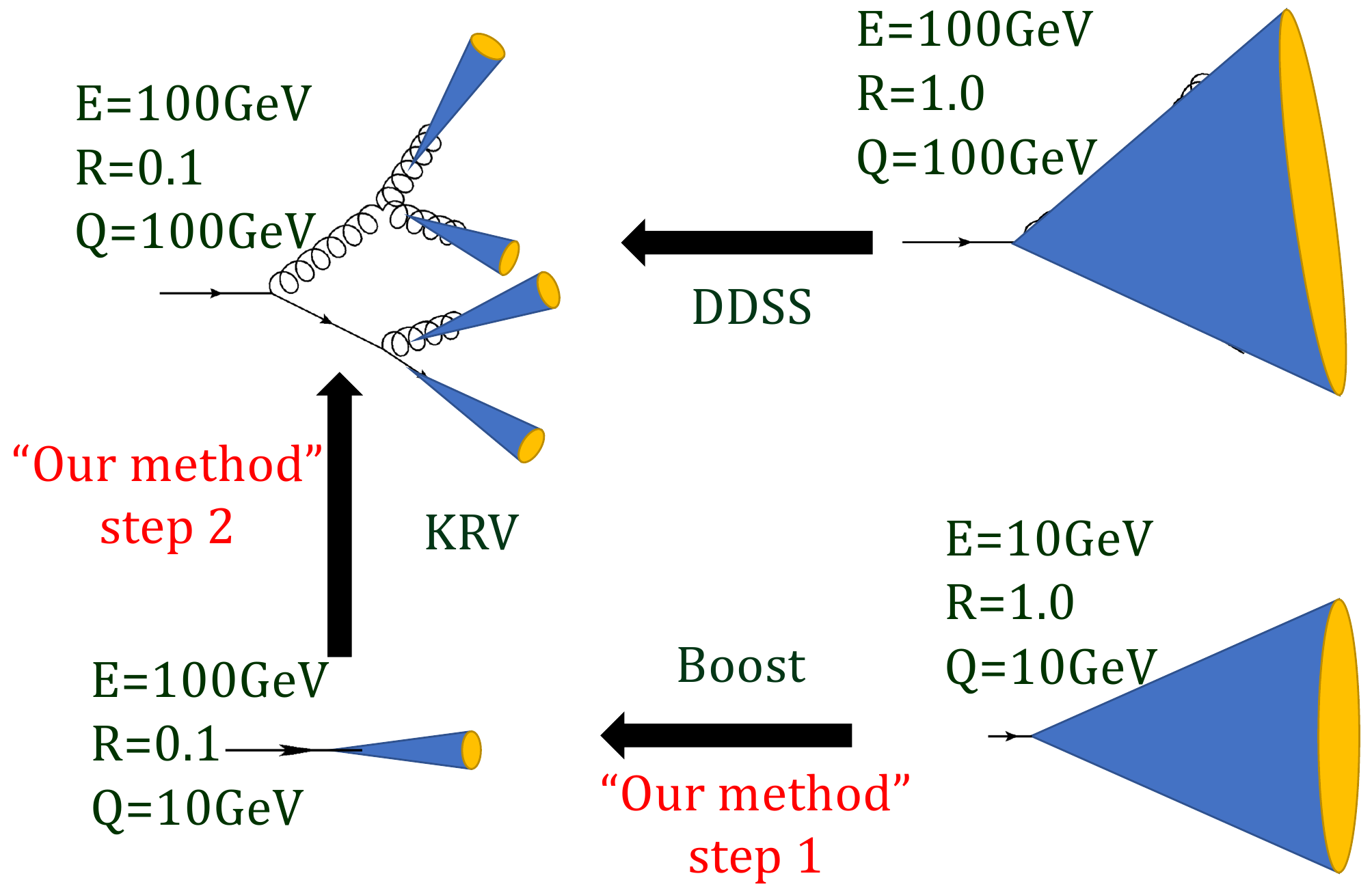}
    \caption{Diagrammatic representation of DDSS approach, KRV approach, and our approach to jet function evolution}
    \label{fig:EvolTree}
\end{wrapfigure}

The evolution method that we use is different from the DDSS and KRV approaches. Fig. \ref{fig:EvolTree} illustrates these three evolution approaches in detail. 

Our evolution approach was then extended to the medium modified evolution of jet functions. We started with the vacuum jet functions at lower energy calculated from PYTHIA simulations and boosted them up to a higher energy so that the jets become significantly narrower,  and the formation time $\tau \approx \frac{E}{Q}$ of the jets are larger so that they form outside the QGP medium. Then we evolved them up to a higher scale using the following medium modified evolution equation,
\begin{equation}
\frac{\partial J_i\left(z, E_J, R, Q\right) \big |_{\xi_i}^{\xi_f}}{\partial \log(Q^2)} = \int_z^1 dy \int_{\xi_i}^{\xi_f} d\xi K _{q^-, Q^2} \left (y, \xi \right ) J_j\left(\frac{z}{y}, E_J, R, Q\right)\bigg |_{\xi_i}^{\xi_f},
\label{SHFF_Med_evol}
\end{equation}
where,
\begin{equation}
K _{q^-, Q^2} \left (y, \xi \right ) = \frac{\alpha _s}{2 \pi} \frac{\tilde{P}_{i \rightarrow j} \left (y \right )}{y Q^2} \hat{q} \left (\xi, x_T \right ) \left [ 2 - 2 \cos \left \{ \frac{ Q^2 \left (\xi - \xi_i \right )}{2 q^- y \left ( 1 - y \right )} \right \} \right ].
\label{Kernel}
\end{equation}

\section{Di-hadron fragmentation function and Soft Drop function}
\vspace{-10pt}
The di-hadron fragmentation function and its evolution are calculated in an extended DGLAP formalism \cite{Majumder:2004wh,Majumder:2004br}. We define a new non-perturbative function, namely "Di-Prong Function" (DPF), as the distribution of the combination of two prongs with momentum fractions $z_1$ and $z_2$ produced in the hadronization of the outgoing hard parton. The operator definition of DPF is given by,
\begin{equation}
\begin{split}
J_q& \left(z_1, z_2, R, Q\approx E_JR\right) = \frac{z^4}{4z_1z_2} \int \frac{dq_\perp^2}{8\left(2\pi\right)}\int \frac{d^4\ell}{(2 \pi)^4} \delta \bigg (z - \frac{p_J^+}{\ell^+}\bigg )\\
& \times \sum_{S_{had}} Tr\bigg [\frac{n\cdot\gamma}{n\cdot p_h} \int d^4x e^{-i\ell\cdot x}   \mel {0}{\psi_q (x)}{p : p_1, p_2, S} \mel {p : p_1, p_2, S}{\bar{\psi}_q (0)}{0}\bigg ].
\end{split}
\label{SDF}
\end{equation}
Even though the operator definition of the DPF is similar to the definition of di-hadron fragmentation function, the evolution equation does not contain the terms responsible for a wider seperation of the two prongs than the radius of the jet, which contains the two prongs \cite{Majumder:2004wh,Majumder:2004br}. However, the angular evolution changes the radius of the jet, and hence partonic splittings can cause a significant effect. 

Similar to the previous procedure for the single jet function, we first calculated the DPFs at lower energy ($E=10$~GeV) from PYTHIA simulations and then boosted them to a higher energy ($E=100$~GeV) and finally evolved them to a higher scale ($Q=100$~GeV) by using the following evolution equation,
 \begin{equation}
\frac{dJ_q \left ( z_1, z_2, R, Q^2\right )}{d\log(Q^2)} = \frac{\alpha_s}{2\pi} \int_z^1 \frac{dy}{y^2} P_{qk} \left(y\right) J_k \left(\frac{z_1}{y}, \frac{z_2}{y}, R, Q^2\right).
\label{SDF_Vac_evol}
\end{equation}

\section{Results and discussion}
\begin{figure}
    \addtolength{\abovecaptionskip}{-3mm}
	\begin{minipage} [l] {0.45\textwidth}
	\centering
	\includegraphics[width=\textwidth]{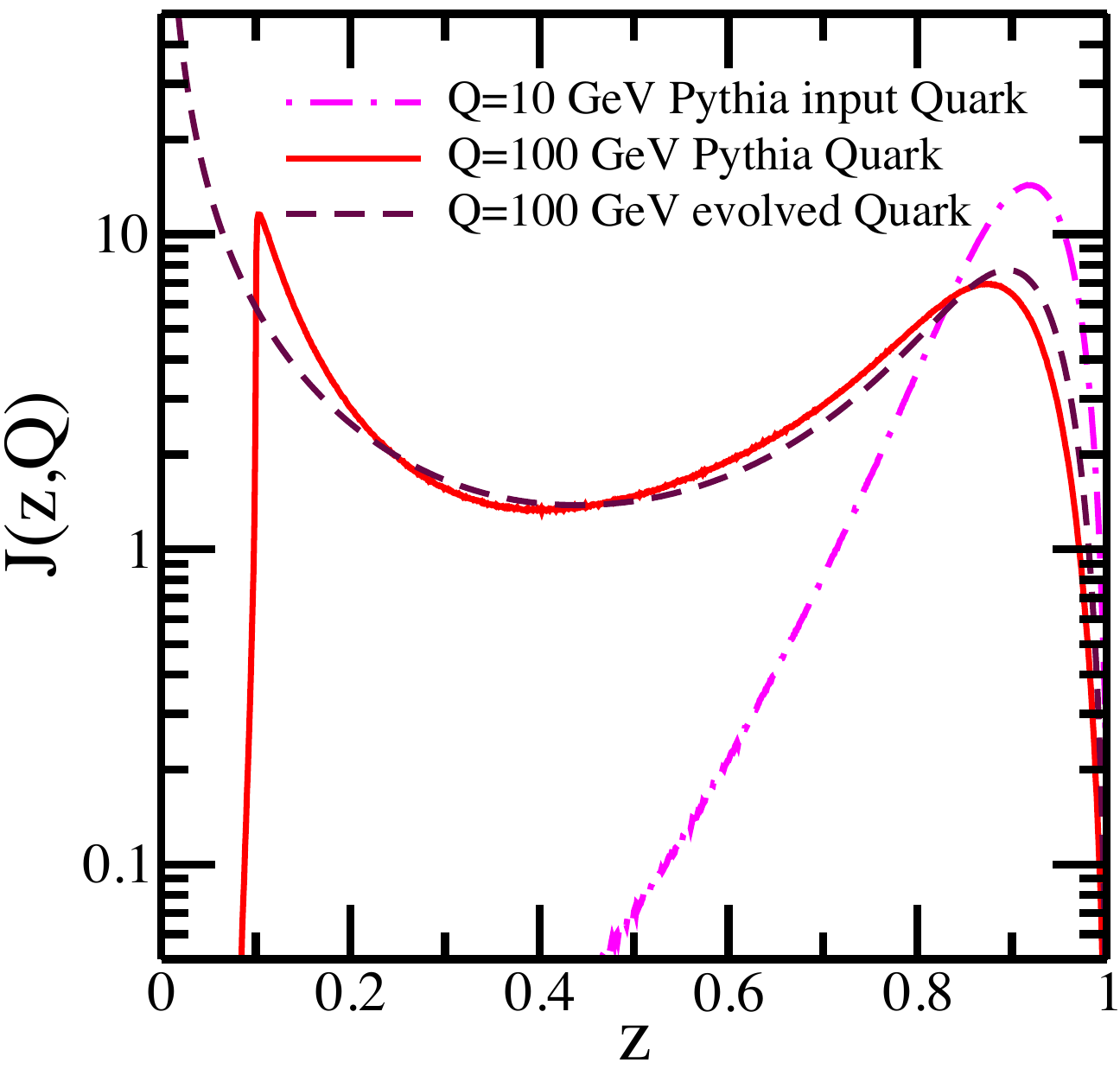}
	\end{minipage}
	\hspace{1cm}
	\begin{minipage} [r] {0.45\textwidth}
	\centering
	\includegraphics[width=\textwidth]{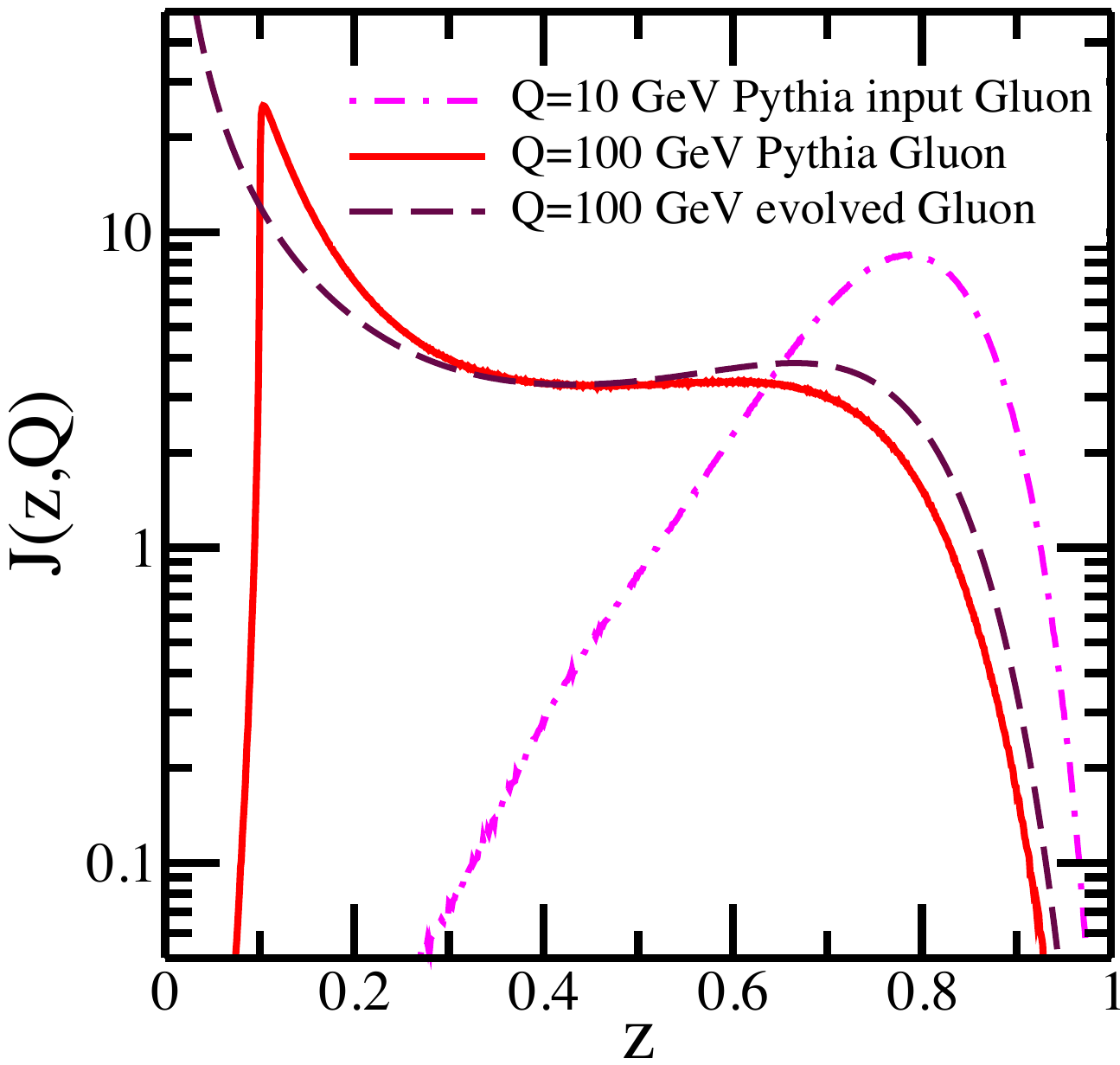}
	\end{minipage}
	\caption{Calculated quark and gluon jet functions at $Q=10$~GeV and $Q=100$~GeV by using PYTHIA simulations and the evolved jet functions at $Q=100$~GeV}
	\label{fig:VacEvol}
\end{figure}

\vspace{-10pt}
The jet functions calculated from PYTHIA simulations at $E=10$~GeV and $E=100$~GeV as well as the evolved jet function at $E=100$~GeV are shown in Fig. \ref{fig:VacEvol}. The evolved jet functions have a reasonable agreement with a direct calculation from PYTHIA at $E=100$~GeV. This validates our evolution approach for the jet function.

\begin{figure}
	\begin{minipage} [b] {0.48\textwidth}
	\centering
    \includegraphics[height=0.88\textwidth]{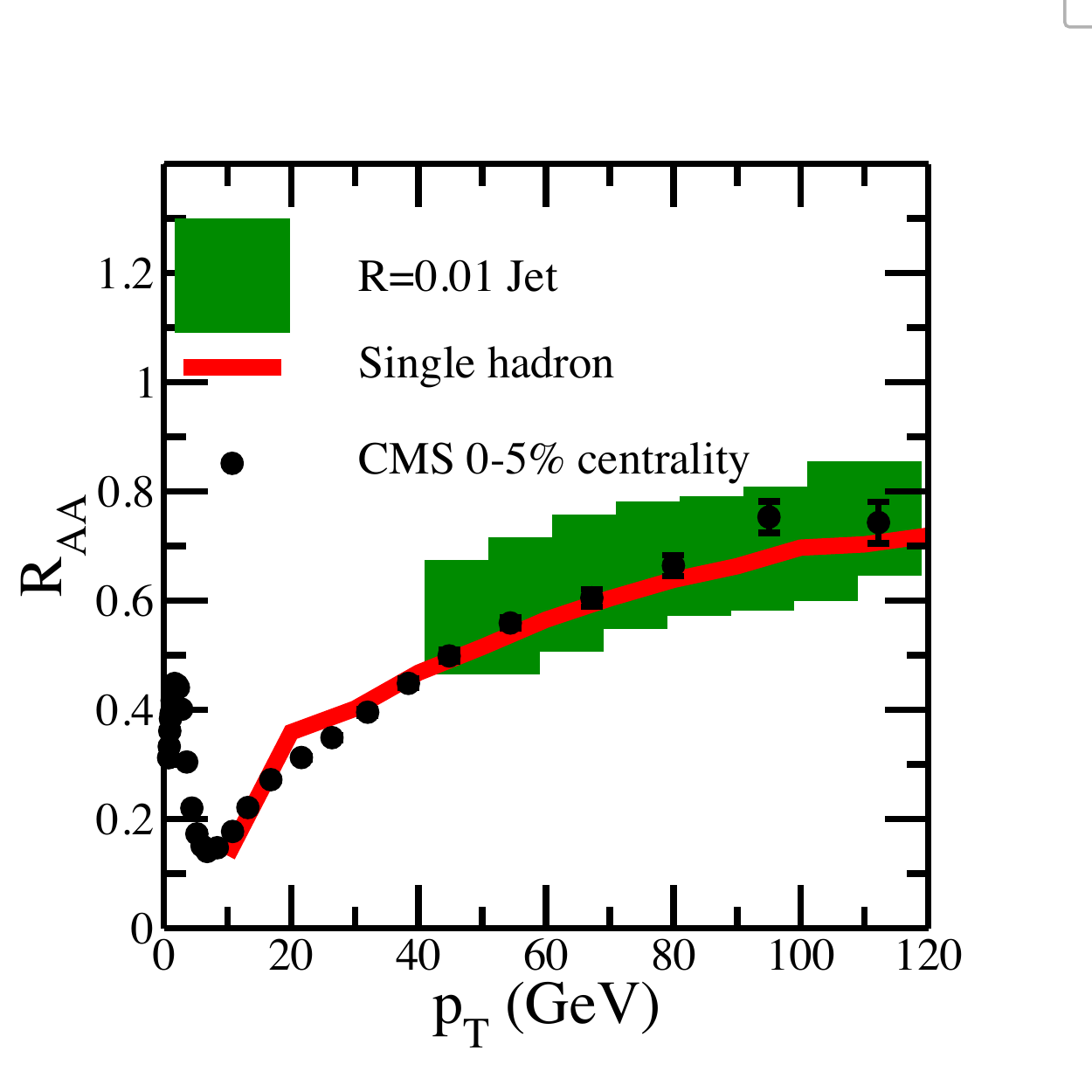}
    \addtolength{\abovecaptionskip}{-3mm}
    \caption{Medium evolution of extremely narrow jets $(R=0.01)$ and leading hadrons in higher-twist formalism compared to data\cite{Khachatryan:2016odn}.}
    \label{fig:MediumJet}
	\end{minipage}
	\hspace{0.2cm}
	\begin{minipage} [b] {0.48\textwidth}
	\centering
    \includegraphics[height=0.88\textwidth]{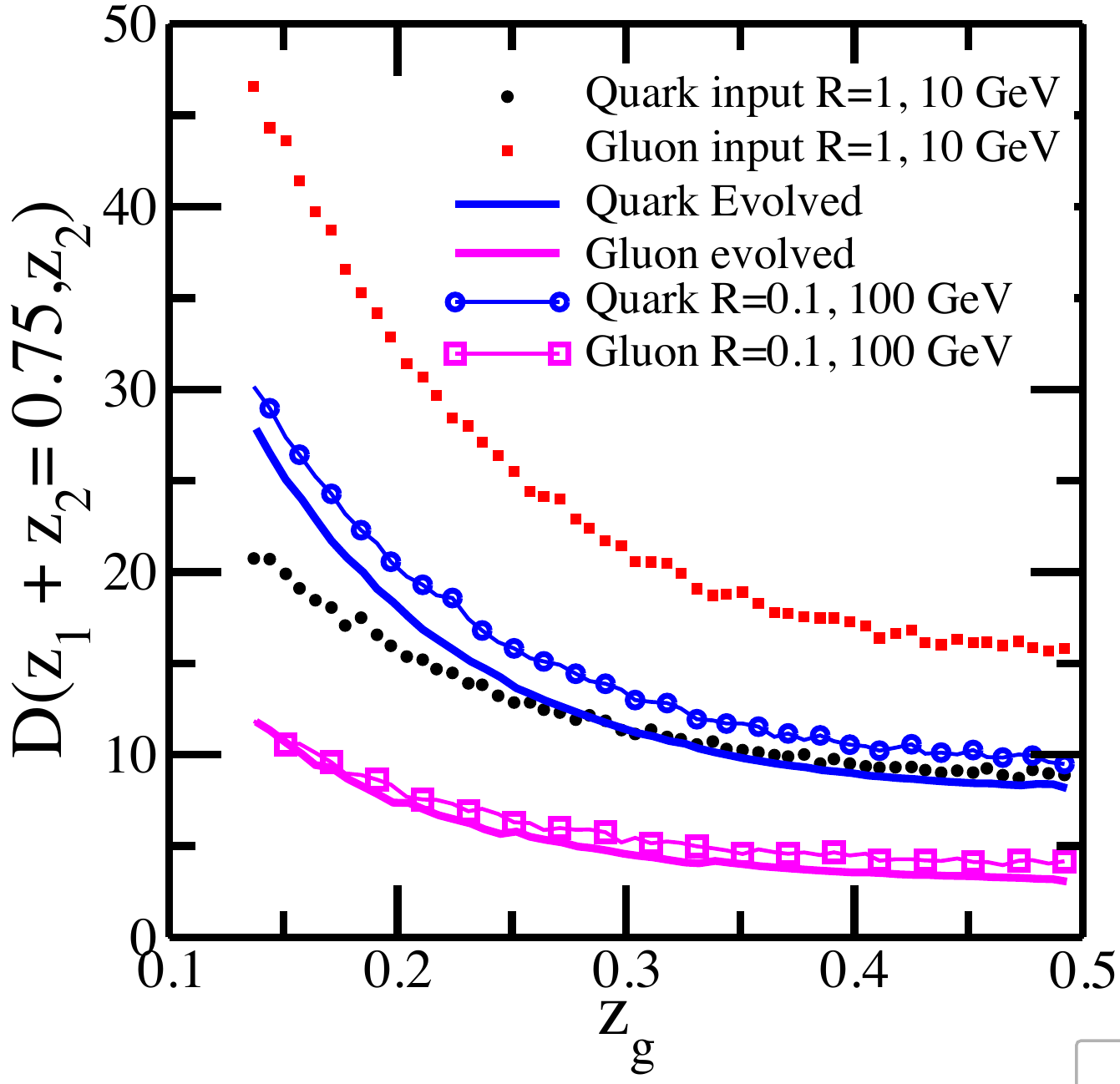}
    \addtolength{\abovecaptionskip}{-3mm}
    \caption{Calculated DPFs at $Q=10$~GeV and $Q=100$~GeV (with $ z=z_1+z_2=0.75$) from PYTHIA  and our evolved DPF at $Q=100$~GeV }
    \label{fig:SDF_Jet}
	\end{minipage}
\end{figure}

In addition to the vacuum evolution, we performed a medium modified evolution within the higher-twist formalism. The nuclear modification factor $R_\mathrm{AA}$ calculated for the extremely narrow jets was in good agreement with the single hadron results from CMS \cite{Khachatryan:2016odn} as well as our direct calculation of single hadron $R_\mathrm{AA}$ by using the higher-twist formalism in Fig. \ref{fig:MediumJet}. Even though this result was expected, this is the first time it has been explicitly demonstrated. When the radius of a jet becomes very small, most jets contain a single hadron, and hence the jet spectrum follows the single hadron spectrum.

Figure \ref{fig:SDF_Jet} shows the calculated DPFs from PYTHIA at $E=10$~GeV and $E=100$~GeV as well as our evolved DPFs at $E=100$~GeV. The evolved DPF results have good agreement with the calculated DPF results in PYTHIA.

\section{Summary and outlook}
\vspace{-10pt}
The vacuum evolution of the single jet function was in good agreement with the PYTHIA simulations. Medium evolution was performed for very narrow jets since the jet should be created outside the QGP medium for the evolution to work. We are working on the angular evolution approach to evolve the jet functions "in angle" to different jet radii inside a medium. Once the angular evolution is calculated, we can use the results from Relativistic Heavy Ion Collider (RHIC) to predict the results from Large Hadron Collider (LHC).

\section{Acknowledgment}
\vspace{-10pt}
This work was supported in part by the U.S. Department of Energy (DOE) under grant number DE-SC0013460 and in part by the National Science Foundation (NSF) under grant number ACI-1550300 within the framework of the JETSCAPE collaboration. CS would like to thank Sara Tipton for the proof reading of the manuscript.

\begin{spacing}{0.0}

\bibliographystyle{JHEP}
\bibliography{Sirimanna_C}

\end{spacing}



\end{document}